\documentclass[format=sigconf,letterpaper]{acmart}
\usepackage{multirow,multicol}
\usepackage{color,tabularx, colortbl,amsmath,bm,arydshln}
\usepackage{enumitem,kantlipsum,arydshln}
\usepackage{booktabs,url,tcolorbox,mathtools} \usepackage{newtxmath}
\usepackage{subfigure}
\usepackage{caption}
\graphicspath{ {figures/} }

\acmConference{Submitted to CIKM 2022}{October 17-21, 2022}{Atlanta}
\acmBooktitle{Proceedings of the 31st ACM International Conference on Information and Knowledge Management}
\copyrightyear{2022}
\acmYear{2022}
\setcopyright{acmlicensed}
\acmPrice{XX.XX}
\acmDOI{XX.XXXX/XXXXXXX.XXXXXXX}
\acmISBN{XXX-X-XXXX-XXXX-X/XX/XX}

\definecolor{Gray}{gray}{0.9}

\begin{document}

\DeclareGraphicsExtensions{.png,.pdf}
\title{ Early Stage Sparse Retrieval with Entity Linking}

\author{	Dahlia Shehata }
\email{dahlia.shehata@uwaterloo.ca		}
\affiliation{%
   \institution{University of Waterloo}   \country{Canada}
}
\additionalaffiliation{
  \institution{Alexandria University, Egypt}
  \country{Egypt} 
}

\author{ Negar Arabzadeh }
\email{narabzad@uwaterloo.ca}
\affiliation{%
   \institution{University of Waterloo}
   \country{Canada}
   }

\author{ Charles L. A. Clarke }
 \email{charles.clarke@uwaterloo.ca}
\affiliation{%
   \institution{University of Waterloo}   \country{Canada}
}
  


\begin{abstract}
Despite the advantages of their low-resource settings, traditional sparse retrievers depend on exact matching approaches between high-dimensional bag-of-words (BoW) representations of both the queries and the collection. As a result, retrieval performance is restricted by semantic discrepancies and vocabulary gaps. On the other hand, transformer-based dense retrievers introduce significant improvements in information retrieval tasks by exploiting low-dimensional contextualized representations of the corpus. While dense retrievers are known for their relative effectiveness, they suffer from lower efficiency and lack of generalization issues, when compared to sparse retrievers. For a lightweight retrieval task, high computational resources and time consumption are major barriers encouraging the renunciation of dense models despite potential gains. In this work, we propose boosting the performance of sparse retrievers by expanding both the queries and the documents with linked entities in two formats for the entity names: 1) explicit and 2) hashed. We employ a zero-shot end-to-end dense entity linking system for entity recognition and disambiguation to augment the corpus. By leveraging the advanced entity linking methods, we believe that the effectiveness gap between sparse and dense retrievers can be narrowed. We conduct our experiments on the MS MARCO passage dataset. Since we are concerned with the early stage retrieval in cascaded ranking architectures of large information retrieval systems, we evaluate our results using recall@1000. Our approach is also capable of retrieving documents for query subsets judged to be particularly difficult in prior work. We further demonstrate that the non-expanded and the expanded runs with both explicit and hashed entities retrieve complementary results. Consequently, we adopt a run fusion approach to maximize the benefits of entity linking.
\end{abstract}


\maketitle

\section{Introduction}
Multi-stage ranking pipelines represent a pivotal transition in Information Retrieval (IR). Early stage retrieval, also known as the recall stage or first stage, aims to find all potentially relevant documents to a query from large collections using inexpensive and efficient ranking models. The retrieved candidate document pool is then forwarded to later reranking stages that employ more complex rankers~---~often neural architectures based on contextualized pre-trained transformers~---~for refinement and pruning. This cascaded ranking pipeline has proved to be highly practical in both academia \cite{optimcascade, Wang2011ACR} and industry \cite{ecommerce, wang2020cold, chen2019behavior}. The objective of the first stage is to efficiently recall a large pool of documents related to the information need. Sparse retrievers such as BM25, with WAND query processing \cite{wand}, have been long prevailed over other retrievers in this stage thanks to their simple logic, inverted index mechanism for large-scale corpora, low requirement of training data, generalization capabilities across different datasets, cost-efficiency, scalability and lower latency. Nonetheless, classical sparse retrievers suffer from the longstanding vocabulary mismatch problem \cite{mismatchavoidance, mismatchir} since they calculate the relevance score by relying on heuristics defined over the exact lexical matching between the queries and the collection. Traditional term-based retrievers use sparse, high-dimensional, bag-of-words (BoW) representations to perform the matching neglecting vocabulary ambiguities and semantic nuances such as synonymy and polysemy. They also fail to capture document semantics because they ignore order dependencies between the terms \cite{semanticmatch}.

Due to these shortcomings, there has been increased research interest in adopting dense retrievers for first-stage ranking. Pre-trained transformer-based dense retrievers offer a significant performance improvement by mapping queries and documents to dense low-dimensional embedding-based contextualized representations to softly match query-document pairs beyond the explicit text surface form \cite{gao2021complementing, khattab2020colbert}. Despite their ability to outperform sparse retrievers, dense retrievers require greater computational resources and a large training corpus, with perhaps hundreds of thousands of labels \cite{msmarco}. In addition to their inability to detect token-level matches \cite{arabzadeh2021predicting}, they also struggle with higher latency issues, lack of generalization \cite{generalization} and lower efficiency compared to classical sparse models. This efficiency-effectiveness tradeoff between sparse and dense retrievers often limits the adoption of the costly dense retrievers to later reranking stages, with a relatively smaller number of retrieved documents, while sparse retrievers such as BM25 are prioritized in the first stage of cascaded ranking systems. Although some approaches suggest leveraging dense retrievers in the early ranking stage, these methods are usually conditioned by hybrid paradigms to maintain efficiency, either by supplementing the sparse retriever by semantic information generated by a dense retriever \cite{gao2021complementing}, interpolating relevance scores of both retrievers \cite{Kuzi2020LeveragingSA}, \cite{lin2020distilling}, \cite{luan2021sparse},
or intelligently selecting the best retriever using a trained classifier \cite{arabzadeh2021predicting}.

Efforts were made to overcome the limitations of sparse retrievers such as query expansion \cite{qexpansion}, document expansion \cite{nogueira2019document}, \cite{Nogueira2019FromDT}, topic models \cite{topicmodel}, translation models \cite{translation} and term dependency models \cite{termdependence}. Nonetheless, advances in this area were relatively slow in contrast with later reranking stages that experienced numerous transformations in the last decade \cite{Cai2021SemanticMF}. In another context, entity linking, an important task of NLP, has been revolutionized in terms of scalability, efficiency and accuracy by recent advances in pre-trained transformer-based architectures. In this work, we aim to leverage the development of entity linking systems to expand the collection with relevant entity names in an attempt to reduce imminent semantic gaps preventing document retrieval for later reranking stages.

Our objective is to prove that in the ``age of muppets'' \cite{zhang-etal-2021-learning-rank}, sparse retrievers can still hold a solid performance boosted by the novel semantic linking systems. Hence, it is possible to shrink the effectiveness gap between sparse and dense retrievers. We conduct our experiments on the MS MARCO passage dataset \cite{msmarco} focusing on the early stage retrieval. Our methods also retrieve relevant query-document matches that were not identified in the non-expanded version of the three so-called Chameleons sets of obstinate queries from MS MARCO~\cite{hardqueries}. Our best-reported results beat standard BM25 results by adopting Reciprocal Rank Fusion (RRF) \cite{rrf} between the original and the entity-aware runs.

Our contributions can be summarized as follows: 1) Wikification of MS MARCO passage dataset using fast end-to-end encoder-based zero-shot entity linking model. 2) Query and document expansion using retrieved entity names in two forms: a) explicit and b) hashed. 3) Run fusion between the non-expanded and the two entity-equipped runs (explicit and hashed forms) to determine the maximum recall@1000 gain achieved by entity linking in comparison with the original BM25 on MS MARCO development set and the three sets of hard queries\cite{hardqueries}.


\section{Related Work}
Leveraging entity linking, with the objective of overcoming difficult matching problems, is not a novel idea in the IR literature. Prior work usually employs end-to-end tools to extract entity mentions from text and link them to their corresponding entity names in a Knowledge Base (KB) before augmenting the representations of the IR corpus.
In addition to the classic BoW representations of queries and documents, there is a rich body of work in literature exploring other representational methods using entities for IR such as: (1) Latent semantic and topic models where the matching takes place when a query and a document share the same set of latent topics. For example, \citet{Liu2015LatentES} introduce a Latent Entity Space (LES) model where queries and documents are projected into a set of latent entities that is used to estimate the document relevance. (2) Bag-of-concepts using multilingual knowledge resources like Wikipedia for cross-language and multilingual IR
\cite{SORG201226, egozi}. (3) Bag-of-entities (BoE) \cite{raviv, docretehsan, Ensan2018AdHR, xiongliu, Gonalves2018ImprovingAH} extracted using automatic linking systems to represent both the queries and the documents. The latter are usually ranked according to the number of occurrences of query entities. Our approach is a combination of both BoW and BoE representations.


Also relevant to our research, we can distinguish the work of Ensan et al. \cite{docretehsan} where the authors introduce entity-based soft matching by proposing the Semantics-Enabled Language Model (SELM) for document retrieval based on the degree of relatedness of the meaning of the query and the documents. They use TAGME \cite{Ferragina2010TAGMEOA}, an entity linking tool, to augment raw text with hyperlinks to corresponding Wikipedia pages, and model queries and documents to sets of semantic concepts connected to each other based on relatedness in an undirected graph. In our work, we use a relatively newer contextualized entity linking model, compared to TAGME, to expand the queries and documents with entity names instead of hyperlinks without further semantic modeling. \citet{Ensan2018AdHR} also explore the interpolation idea between multiple retrievers by building a semantic retrieval framework to increase the relevant results in adhoc keyword-based IR systems. The core of this framework is based on the previously built SELM equipped with two extra semantic analysis configurations, besides TAGME, which are: Explicit semantic analysis (ESA) \cite{esa} and Paragraph2Vec \cite{le2014distributed}. The contributions in \cite{Ensan2019RelevancebasedES} extend the previous two works by addressing research gaps. Although semantic-knowledge-based models depending on entities extracted from KGs were deemed effective for the retrieval performance, these models suffer from topic drift issues. As a result, the authors of this paper introduce the Retrieval through Entity Selection (RES) method by proposing a relevance-based model for entity selection based on pseudo-relevance feedback (PRF) for query expansion and adhoc retrieval. Our work tries to overcome topic drift and matching inconsistency shortcomings in previous research by exploiting advances in recent entity linking models. We have also leveraged PRF as a comparative baseline to our methods.

Closely aligned with our research, \citet{Gonalves2018ImprovingAH} explore the value of entity information for improving adhoc retrieval of feature-based learning-to-rank (LTR) \cite{ltr} search engines.
\citet{duetrepr} demonstrate that a duet term-based and entity-based representations achieves better retrieval results compared to the standalone BoW or BoE approaches.
As shown in \cite{Gonalves2018ImprovingAH}, adopting LTR methods for semantic retrieval is an active research direction where semantic information is integrated into ranking models. In this context, several research works leverage named entities \cite{nattiya} or semi-structured meta-data \cite{esdrank} as additional features for the purpose of learning ranking architectures.
The empirical study in \cite{Ensan2017AnES} examines the effectiveness of joining document neural embedding features with entity embeddings using LTR methods. Although this line of prior works adopts the same combined BoW and BoE approach, they use entities as extra features. In contrast, we expand BOW representations with term-based entities. To the best of our knowledge, our methods have not been previously experimented in literature.


\begin{figure*}[ht]
  \centering
  \subfigure[Dev set.]{\includegraphics[scale=0.149]{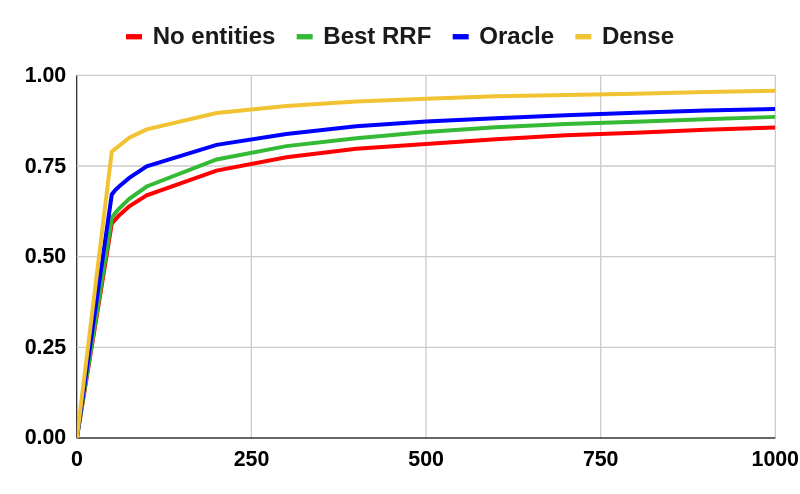}}\quad
  \subfigure[Hard set.]{\includegraphics[scale=0.149]{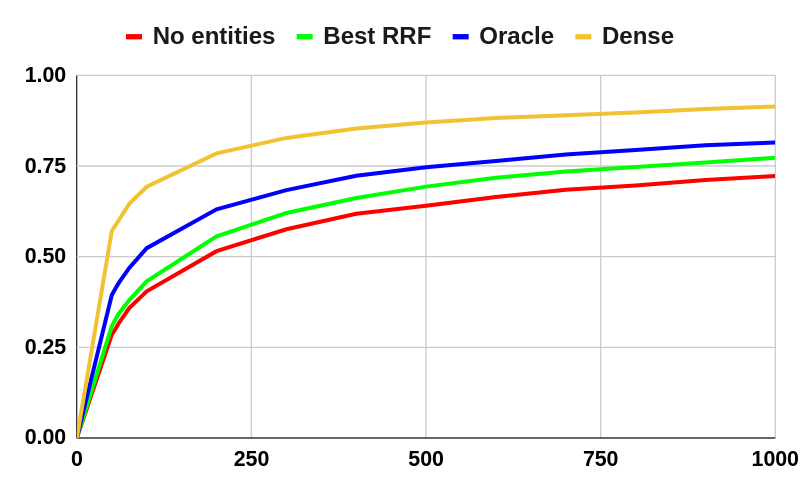}}\quad
  \subfigure[Harder set.]{\includegraphics[scale=0.149]{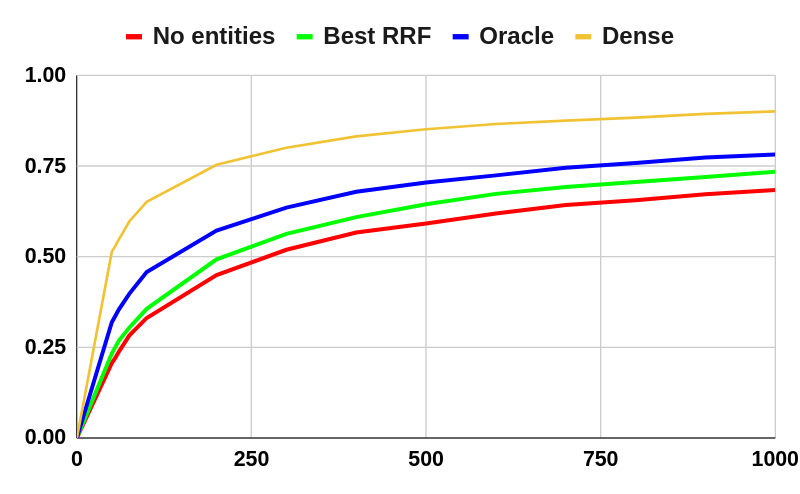}}\quad
  \subfigure[Hardest set.]{\includegraphics[scale=0.149]{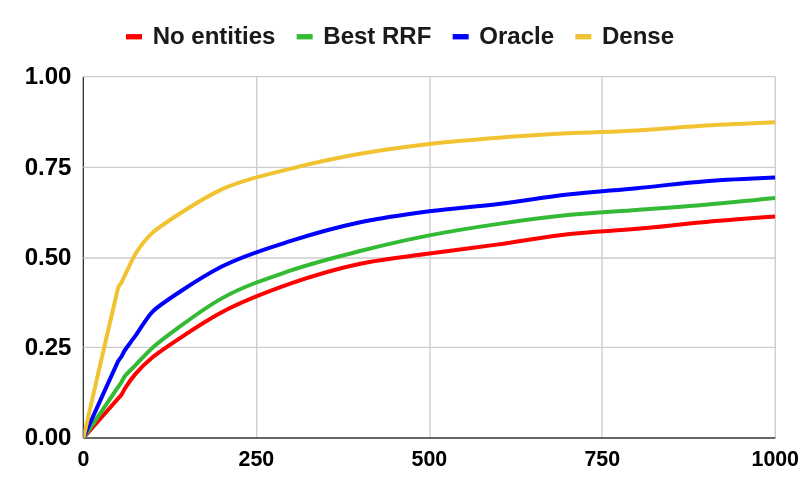}}\quad
  \vspace{-1em}

  \caption{Recall curves of different query sets.
  The x-axis shows the cutoffs, and the y-axis is the corresponding recall value.
  Although the ANCE (yellow) curves are the highest, entity linking introduces significant improvement to the classical BM25 without entities (red).  The oracles (blue) show the maximum possible gain achieved by entity linking. The best RRF of run combinations are shown in green.}
  \vspace{-1em}
    \label{fig:recall_curves}

\end{figure*}


\section{Methods}
In order to create the best pool for reranking, we aim to maximize the recall for the first stage retrieval. Our experiments are conducted on the Microsoft MAchine Reading Comprehension (MS MARCO) passage collection\footnote{https://microsoft.github.io/msmarco/}. 
This collection comprises 8.8 million passages, along with over 500k pairs of query and judged-relevant passages for training purposes.
For less than 10\% of the queries, there are multiple judged relevant passages per query. 
The “MS MARCO Small Development Set” includes 6,980 queries dedicated to development and validation. There is also a test set with private (not publicly available) relevance judgments for leaderboard purposes. In our work, we use the training and the development sets. We refer to the small development set as ``Dev'' set for conciseness.

We further test our techniques on the MS MARCO Chameleons sets of obstinate queries \cite{hardqueries}.
These query sets are subsets of the MS MARCO queries that are difficult for state-of-the-art rankers to satisfy and they showed extremely poor performance.
They do not experience any performance improvement regardless of the underlying ranker, i.e. the overall improvement reported by a ranker always results from another subset of queries.
The MS MARCO Chameleons set consists of three main sets: 1) Veiled Chameleon (or ``Hard'' set) comprises 3,119 hard queries that are common among the worst 50\% of the performing queries of at least four rankers. 2) Pygmy Chameleon (or ``Harder'' set) includes 2,473 hard queries that are common between the worst 50\% of queries of at least five rankers. 3) Lesser Chameleon (or ``Hardest'' set) comprises 1,693 queries judged as the hardest by six rankers. We demonstrate that collection expansion with linked entities helps rankers to discover a higher percentage of the hard queries in these sets.

\subsection{Entity Linking}
We use ELQ \cite{elq}, a fast end-to-end entity linking system. Although no entity disambiguation system can fit all datasets, ELQ achieves state-of-the-art performance compared to other end-to-end tools \cite{shen2021entity}. ELQ performs both entity recognition and disambiguation in one pass. The system determines each entity mention boundaries in a given question and the corresponding Wikipedia entity using a BERT-based bi-encoder. First, the entity encoder embeds every Wikipedia entity using its short description. Then, the question encoder calculates token-level embeddings for the input question. These two embeddings are finally leveraged in mention boundary detection and disambiguation by computing their inner product.

ELQ is built on BLINK \cite{blink}, a two-stage zero-shot BERT-based linking algorithm. Unlike ELQ, BLINK only performs entity disambiguation using pre-specified entity mention boundaries in the input. The first stage uses a biencoder consisting of two BERT transformers to independently encode mention context and entity descriptions into dense vectors. Entity candidates are scored as vector dot products. The retrieved candidates are then forwarded to the second stage for reranking. The latter takes place using a cross-encoder that concatenates and encodes the mention and the text in one transformer.

We performed entity linking on MS MARCO for both the queries and the passages. Since ELQ is originally dedicated to short-length questions, we adopted an overlapping sliding window approach to extract entities from longer passages using a context window size of 128 tokens and an overlap stride of 42, so that the window overlaps by $1/3$ of the text length. 
This overlap ensures that for each subpassage, the context is taken into account from both sides. The entity set of the whole passage is later deduplicated. We also retained the default parameter settings ($threshold=4.5$, $num\_cand\_mentions=10$, $num\_cand\_entities=10$) recommended by ELQ.

\subsection{Corpus Expansion}
We expand both the queries and the passages with a single instance of each retrieved entity name. We attempt collection augmentation with entities using two forms: explicit word form and MD5 hashed form.
The intuition behind our decision to experiment with MD5 hashed entities is to provide consistent representations of multi-word terms, hence avoiding partial or wrong matching between a query and a non-relevant passage. 
In addition to expansion using one copy of each entity name, we have also experimented with weighted expansion reflecting the number of entity mention occurrences in the text, and expansion with a constant factor.
As a sparse retriever, we employ BM25 as implemented by the open-source Anserini system \cite{Yang2017AnseriniET}, which provides state-of-the-art performance for sparse retrievers.
The Anserini\footnote{\url{https://github.com/castorini/anserini}} implementation of BM25 has been widely adopted as the first stage retriever in many multi-stage ranking stacks~\cite{hofsttter2020local, hofsttter2020interpretable, nogueira2019multistage}.
We use the dense retriever ANCE ~\cite{xiong2020approximate} as a basis for our comparisons since it is a well-established contrastive representation learning mechanism for dense retrieval using an asynchronously updated Approximate Nearest Neighbor (ANN) index.
The following example shows one of the hardest queries that vanilla BM25 fails to retrieve, but is correctly identified thanks to entity linking. The entities are shown in their explicit form in red.

\textbf{Query:} \textit{who are in the eagles \textcolor{red}{\underline{Eagles (band)}}}

\textbf{Passage:} \textit{Who are the original members of The Eagles rock band? Glenn Frey, Don Henley, Bernie Leadon and Randy Meisner are the four original members who formed The Eagles rock band in Los Angeles, California in 1971.     \textcolor{red}{\underline{Glenn Frey} \underline{Don Henley} \underline{Randy Meisner} \underline{Bernie Leadon} \underline{Los Angeles} \underline{California} \underline{Eagles (band)}}}




\subsection{Run Combination}
In order to estimate the maximum recall gain that can be achieved by entity linking, we generate hypothetical oracle runs for each query set by selecting the run with the highest passage rank for each query. If all the three BM25 runs under consideration (with no entities, with entities, with hashed-entities) do not include the passage required by the qrel set for a given query, the run selection is performed arbitrarily, since the recall will always be zero in any case.  For cases where there are multiple judged relevant passages per query, we prioritize the judged passage with the highest rank across all runs. In order to reduce the margin between the individual runs and the oracle results, we experimented with Reciprocal Rank Fusion (RRF) \cite{rrf} for all the combinations of the three mentioned runs. RRF combines the passage rankings from multiple runs by sorting the passages according to a simple scoring formula achieving better results than any individual run.

\begin{table}[!t] 
\Small
\caption{Recall@1000 of the 4 query sets. PRF stands for Pseudo-Relevance Feedback. 
}
\vspace{-1em}
\begin{tabular}{l|cccc} 
&
\multicolumn{3}{c} \textbf{\textbf{Query Set Type}}
 \\ [0.5ex]
\hline
\textbf{Run type}&\textbf{ Dev}&\textbf{Hard}&\textbf{Harder}&\textbf{Hardest}
\\[0.5ex] 
\hline 
No entities&0.8573&0.7234&0.6849&0.6136 \\


Hashed entities&0.8479&0.7146&0.6727&0.5995\\

Entities&0.8682&0.7467&0.7079&0.6389 \\


No entities/ Hashed entities  RRF&0.8780&0.7591&0.7195&0.6471 \\

Hashed entities/ Entities RRF&0.8784&0.7599&0.7196&0.6498 \\

No entities/ Entities RRF&0.8844&0.7695&0.7323	&0.6625 \\

No entities/ Entities/ Hashed RRF&\textbf{0.8868$^{}$}&\textbf{0.7738$^{}$}&\textbf{0.7353$^{}$}&\textbf{0.6650$^{}$}\\
 
\arrayrulecolor{gray}\hline

No entities + PRF &0.8759&0.7622&0.7272&0.6674\\

\arrayrulecolor{gray}\hline

Oracle&0.9087&0.8159&0.7827&0.7220\\

\arrayrulecolor{gray}\hline

Dense&0.9587&0.9152&0.9022& 0.8753 \\

\arrayrulecolor{black}\hline

\end{tabular} 
\vspace{-1em}
\label{tab:recall}
\end{table}


\section{Experiments}

\subsection{Experimental Setup}
We conduct our experiments using a cluster of Intel E5-2683 v4 Broadwell - 2.1Ghz for entity inference and Anserini-related experiments such as the run generation using the training query set. 

\subsection{Results}

We have experimented with corpus expansion using three approaches: 1) A single copy of the entity name. 2) A constant number of copies of each entity name such as 3 and 5. 3) Weighted expansion according to the number of entity mention occurrences. However, using a single entity term for each detected mention gives the best results. In fact, we have found that the expansion with multiple copies of the same entity is inversely proportional to the recall performance, i.e factor 5 gives worse results than 3.

Since we are concerned with maximizing the performance of the first stage retrieval for later reranking, we evaluate our results using recall@1000. We use BM25 tuned hyperparamaters ($k_{1}=0.82, b=0.68$) that are optimized for recall@1000 on the MS MARCO dataset in Anserini. Although, we have attempted tuning these parameters on the entity-equipped dataset with both versions: explicit and hashed, there were no considerable changes in the final hyperparamater values.

 As shown in Table \ref{tab:recall}, entity-equipped runs, using the entity explicit format (row 3), gave better recall performance compared to the original BM25 runs with no entities (row 1) across the MS MARCO Dev set and the three sets of obstinate queries.  The improvement gain is observed even without the adoption of further run fusion approaches. This result demonstrates that semantic expansion helps rankers disambiguate the hard queries. To further investigate the entity effect, we have experimented the performance with the hashed version. We can observe that the individual hashed-entity-equipped runs (row 2) have worse recall results than the original ones (row 1). Nonetheless, the pairwise reciprocal rank fusion between the original runs and those with the hashed entities (row 4) outperforms the three individual runs: original, with hashed entities, with entities  (i.e. the first 3 rows) for all types of queries. This could be because the runs expanded with hashed entities fetch complementary results that are not retrieved by BM25 using the non-expanded dataset. Nonetheless, further investigation is still required to hypothesize the bad performance of the individual hashed-entity-equipped runs.

The best recall results are achieved using the RRF of the three runs with a statistically significant performance improvement of 3.44\%, 6.97\%, 7.36\% and 8.38\% for the Dev (p-value < 0.05), Hard, Harder and Hardest query sets respectively.  The statistical significance of the results was verified using paired t-test. 
We have also noticed that the hashed entity-equipped run contributes to the overall gain by only a small factor. This can be clearly seen when comparing the results of the pairwise RFF of the no-entity and the entity-aware runs (row 6), and the RRF of the three runs (row 7). The hypothetical oracle runs exceed the best-achieved results with percentages of 2.47\%, 5.44\%, 6.45\% and 8.57\% for the very same sets demonstrating that a room for improvement remains available with the right run combination or selection strategy. 
The latter is worth exploring in a related future work.
In addition to the Oracle (row 9) and the ANCE (row 10) results that we use as a comparative reference, we have also investigated the pseudo-relevance feedback (PRF) effect on the non-expanded MS MARCO (row 8). Although PRF causes a significant gain with a recall@1000 of 0.8759 compared to 0.8573 on the Dev set, we refrain from including the costly PRF in our entity-related experiments. It is interesting though to examine PRF effect on the entity-aware dataset. We can also see that RRF of the three runs (row 7) still outperforms the BM25+PRF non-expanded run (row 8) across the Dev, Hard and Harder sets. However, PRF results are still slightly higher for the Hardest set.



Figure \ref{fig:recall_curves} illustrates the effectiveness differences between the recall curves of four main runs: the original BM25 with no entities (red), the best combination of no-entity and entity-aware BM25 runs that is achieved by RRF for a given query set (green), the hypothetical oracle (blue) and the ANCE run (yellow). The curves cover four query sets: the Dev, Hard, Harder and Hardest sets. The x-axis represents the different cutoffs while the y-axis shows the corresponding recall results. As demonstrated by the yellow curves in Figure \ref{fig:recall_curves}, and also in the dense results of Table \ref{tab:recall}, ANCE retrieval still outperforms all BM25-dependent retrieval by a significant margin. Nonetheless, we observe that the effectiveness difference between BM25 and ANCE has considerably decreased with the help of semantic linking. The oracle curves suggest that an additional performance improvement is still possible by taking advantage of linked entities, further reducing the recall gap between sparse and dense retrievers. 


\section{Conclusion}
In contrast with dense retrievers, sparse retrievers offer higher efficiency benefits at the expense of semantic comprehension. In an attempt to bridge the gap, we propose leveraging recent advances in entity linking to expand IR collections, hence reducing vocabulary discrepancies. We focus on boosting the retrieval recall in the first retrieval stage 
using BM25.
Our best results are achieved by RRF between different run combinations. Through comparative evaluation, we prove that our approach enhances the performance of traditional sparse retrieval, with additional potential for improvement.

\newpage
\bibliographystyle{ACM-Reference-Format}
\balance
\bibliography{acmart} 

\end{document}